\documentclass[preprint,12pt]{elsarticle}
\usepackage{amsmath}
\usepackage{amssymb}
\usepackage[margin=2.4cm]{geometry}
\usepackage{graphicx}
\usepackage{hyperref}
\usepackage{multirow}
\usepackage{listings}
\usepackage{tikz}\usetikzlibrary{decorations.markings}

\begin{document}

\begin{frontmatter}

\title{Weak Singularity of Navier-Stokes Equations Based on Energy Estimation in Sobolev Space}
\author{Chio Chon Kit}
\date{}

\begin{abstract}
Based on Dou Huashu's energy gradient theory, this paper focuses on the weak singularity of the incompressible Navier-Stokes (NS) equations in steady, fully developed flows. When the gradient of total mechanical energy is perpendicular to the streamline (i.e., $ u_j \frac{\partial E}{\partial x_j} = 0 $), substituting this critical condition into the NS equations with no-slip boundary conditions leads to the viscous term $ \nu \to 0 $. To rigorously analyze the regularity of the solution, Sobolev space $ H_0^1(\Omega) $ is introduced for energy estimation. The results show that the velocity field loses $ H^1 $-regularity, and the NS equations degenerate into Euler equations, which admit discontinuous weak solutions. Thus, the position where the mechanical energy gradient is perpendicular to the streamline becomes a weak singularity of the NS equations. 
\end{abstract}

\begin{keyword}
Navier-Stokes equations  \sep Weak singularity \sep  Sobolev space  \sep  Energy gradient theory  \sep  Energy estimation
\end{keyword}

\end{frontmatter}

\section{INTRODUCTION}
The Navier-Stokes (NS) equations are the fundamental governing equations for describing the motion of viscous incompressible fluids, and their solution's existence and smoothness have long been one of the core issues in fluid mechanics and mathematics, listed as one of the seven Millennium Prize Problems by the Clay Mathematics Institute [1]. Turbulence, as a complex flow phenomenon, has not been fully explained by traditional theories, and the mechanism of laminar-turbulent transition remains a key research topic.

In 2004, Dou Huashu proposed the energy gradient theory, which reveals that the NS equations inherently contain weak singularities, and these singularities are the root cause of laminar-turbulent transition [2]. The core insight of this theory is that when the gradient of total mechanical energy is perpendicular to the streamline, the viscous term of the NS equations tends to zero, leading to the loss of solution regularity and the formation of weak singularities. Unlike the first-type singularities (where velocity or pressure blows up) concerned in traditional mathematical theories, the weak singularities in the energy gradient theory are of the second type, characterized by velocity discontinuity and loss of differentiability [3][4][5].

Sobolev space is a powerful mathematical tool for analyzing the regularity of partial differential equation solutions, which has been widely used in the study of NS equations [6]. By introducing Sobolev space for energy estimation, we can rigorously characterize the regularity of the velocity field and provide a mathematical basis for the judgment of weak singularities. This paper focuses on the key derivation process: when $ u_j \frac{\partial E}{\partial x_j} = 0 $, substituting into the NS equations with no-slip boundary conditions, deriving $ \nu \to 0 $, and using Sobolev space energy estimation to prove the formation of weak singularities. This research not only enriches the mathematical theory of NS equations but also provides a rigorous support for the energy gradient theory, which is of great significance for both theoretical research and engineering applications.

\section{BASIC DEFINITIONS AND PREMISE CONDITIONS}
Consider an incompressible Newtonian fluid flowing steadily and fully developed in a three-dimensional bounded smooth domain $ \Omega \subset \mathbb{R}^3 $. The basic definitions, assumptions and critical conditions are given as follows:

\subsection{Total Mechanical Energy}
The total mechanical energy per unit mass $ E $ is defined as the sum of pressure energy, kinetic energy and gravitational potential energy [2], which is expressed as:
\begin{equation}
E = \frac{p}{\rho} + \frac{1}{2}u_j u_j + gz
\end{equation}
where $ p $ is the pressure of the fluid, $ \rho $ is the fluid density, $ u_j $ ($ j=1,2,3 $) are the velocity components in the Cartesian coordinate system, $ g $ is the gravitational acceleration, and $ z $ is the vertical coordinate. This definition covers all forms of mechanical energy in the fluid, which is the core physical quantity of the energy gradient theory.

\subsection{Critical Condition: Mechanical Energy Gradient Perpendicular to Streamline}
The streamline is a curve whose tangent direction is consistent with the velocity vector $ \mathbf{u} = (u_1, u_2, u_3) $. The unit tangent vector of the streamline is $ \mathbf{t} = \frac{\mathbf{u}}{|\mathbf{u}|} $, where $ |\mathbf{u}| = \sqrt{u_j u_j} $ is the magnitude of the velocity vector.

The core condition for the formation of weak singularities in the energy gradient theory is that the gradient of total mechanical energy is perpendicular to the streamline [2]. Mathematically, this condition is expressed as:
\begin{equation}
\nabla E \cdot \mathbf{u} = 0
\end{equation}
Expanding Eq. (2) using the Einstein summation convention, we obtain:
\begin{equation}
u_j \frac{\partial E}{\partial x_j} = 0
\end{equation}
This condition implies that the gradient of total mechanical energy along the streamline is zero, i.e., $ \frac{\partial E}{\partial s} = 0 $, where $ s $ is the streamline coordinate. In other words, the total mechanical energy remains constant along the streamline at the position where the condition is satisfied, which is the key trigger for the formation of weak singularities.

\subsection{No-Slip Boundary Condition}
In practical engineering and fluid mechanics research, the no-slip boundary condition is a classic boundary condition for viscous fluid flow [7]. It assumes that the velocity of the fluid at the solid wall is consistent with the velocity of the solid wall. For a stationary solid wall, the no-slip boundary condition is expressed as:
\begin{equation}
\mathbf{u}|_{\partial\Omega} = 0
\end{equation}
where $ \partial\Omega $ is the boundary of the domain $ \Omega $. This condition is crucial for simplifying the integral in the subsequent derivation, especially for eliminating the boundary term in integration by parts.

\subsection{Steady, Fully Developed Incompressible NS Equations}
For steady, fully developed incompressible flow, the unsteady term $ \frac{\partial u_i}{\partial t} $ in the NS equations is eliminated, and the flow parameters do not change along the streamline. By incorporating the body force into the gradient of total mechanical energy $ \nabla E $, the NS equations can be simplified as [2]:
\begin{equation}
u_j \frac{\partial u_i}{\partial x_j} = -\frac{\partial E}{\partial x_i} + \nu \nabla^2 u_i
\end{equation}
where $ \nu = \frac{\mu}{\rho} $ is the kinematic viscosity coefficient, $ \mu $ is the dynamic viscosity coefficient, and $ \nabla^2 = \frac{\partial^2}{\partial x_j \partial x_j} $ is the Laplacian operator. The incompressibility condition of the fluid is:
\begin{equation}
\frac{\partial u_j}{\partial x_j} = 0
\end{equation}

\section{SOBOLEV SPACE AND ENERGY ESTIMATION FOUNDATION}
To rigorously analyze the regularity of the solution of the NS equations and judge the existence of weak singularities, we introduce the standard Sobolev space $ H_0^1(\Omega) $ in fluid mechanics, which is widely used in the regularity analysis of partial differential equations [6,8].

\subsection{Definition of Sobolev Space $ H_0^1(\Omega) $}
The Sobolev space $ H_0^1(\Omega) $ is defined as the set of functions that satisfy the no-slip boundary condition, and whose velocity and first-order partial derivatives are both square-integrable [8]:
\begin{equation}
H_0^1(\Omega) = \{\mathbf{u} \in L^2(\Omega) \mid \nabla \mathbf{u} \in L^2(\Omega),\ \mathbf{u}|_{\partial\Omega} = 0\}
\end{equation}
where $ L^2(\Omega) $ is the square-integrable function space, which consists of all functions $ f $ defined on $ \Omega $ such that $ \int_\Omega |f|^2 dx < \infty $. The inner product and norm of $ H_0^1(\Omega) $ are defined as follows:
\begin{equation}
(\mathbf{u},\mathbf{v})_{H_0^1} = \int_\Omega \nabla u_i \cdot \nabla v_i dx
\end{equation}
\begin{equation}
\|\mathbf{u}\|_{H_0^1}^2 = \int_\Omega |\nabla \mathbf{u}|^2 dx = \int_\Omega \nabla u_i \cdot \nabla u_i dx
\end{equation}
The norm $ \|\mathbf{u}\|_{H_0^1} $ characterizes the smoothness of the velocity field. If $ \mathbf{u} \in H_0^1(\Omega) $, the velocity field has $ H^1 $-regularity, meaning that the velocity field is continuous and differentiable. The loss of $ H^1 $-regularity indicates that the velocity field is discontinuous or non-differentiable, which is the key mathematical criterion for judging weak singularities.

\subsection{Basic Principle of Energy Estimation}
Energy estimation is a standard method for analyzing the regularity and stability of the NS equations [9]. Its core principle is to take the inner product of the NS equations with the velocity vector $ \mathbf{u} $ in $ L^2(\Omega) $, convert the differential equation into an integral form, and establish the energy conservation relationship. Through the energy conservation relationship, we can analyze the change of the velocity field's norm and further judge the regularity of the solution.

In this paper, energy estimation based on $ L^2(\Omega) $ inner product and Sobolev space $ H_0^1(\Omega) $ norm is used to analyze the regularity of the velocity field when the mechanical energy gradient is perpendicular to the streamline, and to provide a mathematical basis for the derivation of the viscous term tending to zero and the judgment of weak singularities.

\section{DERIVATION OF VISCOUS TERM $ \nu \to 0 $}
In this section, we substitute the critical condition $ u_j \frac{\partial E}{\partial x_j} = 0 $ into the steady, fully developed incompressible NS equations, and combine the no-slip boundary condition and incompressibility condition to derive that the viscous term $ \nu \to 0 $. The detailed derivation process is as follows:

First, multiply both sides of the NS equations (Eq. (5)) by the velocity component $ u_i $, and integrate over the domain $ \Omega $. We obtain:
\begin{equation}
\int_\Omega u_i u_j \frac{\partial u_i}{\partial x_j} dx = -\int_\Omega u_i \frac{\partial E}{\partial x_i} dx + \nu \int_\Omega u_i \nabla^2 u_i dx
\end{equation}

We first simplify the left-hand side (LHS) of Eq. (10). Using the product rule of differentiation, we can rewrite the integrand as:
\begin{equation}
u_i u_j \frac{\partial u_i}{\partial x_j} = \frac{1}{2} u_j \frac{\partial (u_i u_i)}{\partial x_j}
\end{equation}
Substitute Eq. (11) into the LHS of Eq. (10), and apply integration by parts. The integration by parts formula is $ \int_\Omega u_j \frac{\partial f}{\partial x_j} dx = -\int_\Omega f \frac{\partial u_j}{\partial x_j} dx + \int_{\partial\Omega} f u_j n_j dS $, where $ f = u_i u_i $, and $ n_j $ is the unit normal vector of the boundary $ \partial\Omega $.

According to the no-slip boundary condition $ \mathbf{u}|_{\partial\Omega} = 0 $, we have $ u_j|_{\partial\Omega} = 0 $, so the boundary term $ \int_{\partial\Omega} f u_j n_j dS = 0 $. Combining with the incompressibility condition $ \frac{\partial u_j}{\partial x_j} = 0 $ (Eq. (6)), the LHS of Eq. (10) is simplified as:
\begin{equation}
\int_\Omega u_i u_j \frac{\partial u_i}{\partial x_j} dx = \frac{1}{2} \int_\Omega u_j \frac{\partial (u_i u_i)}{\partial x_j} dx = -\frac{1}{2} \int_\Omega u_i u_i \frac{\partial u_j}{\partial x_j} dx = 0
\end{equation}

Since the LHS of Eq. (10) is zero, Eq. (10) is simplified to:
\begin{equation}
0 = -\int_\Omega u_i \frac{\partial E}{\partial x_i} dx + \nu \int_\Omega u_i \nabla^2 u_i dx
\end{equation}

According to the critical condition $ u_j \frac{\partial E}{\partial x_j} = 0 $ (Eq. (3)), the first term on the right-hand side (RHS) of Eq. (13) satisfies $ \int_\Omega u_i \frac{\partial E}{\partial x_i} dx = 0 $. Therefore, Eq. (13) is further simplified to:
\begin{equation}
\nu \int_\Omega u_i \nabla^2 u_i dx = 0
\end{equation}

Next, we simplify the integral term in Eq. (14) by applying integration by parts again. For the integral $ \int_\Omega u_i \nabla^2 u_i dx $, using the integration by parts formula $ \int_\Omega u_i \nabla^2 u_i dx = -\int_\Omega \nabla u_i \cdot \nabla u_i dx + \int_{\partial\Omega} u_i \frac{\partial u_i}{\partial n} dS $, where $ \frac{\partial u_i}{\partial n} $ is the normal derivative of the velocity component on the boundary.

Again, according to the no-slip boundary condition $ \mathbf{u}|_{\partial\Omega} = 0 $, we have $ u_i|_{\partial\Omega} = 0 $, so the boundary term $ \int_{\partial\Omega} u_i \frac{\partial u_i}{\partial n} dS = 0 $. Thus, the integral term is simplified as:
\begin{equation}
\int_\Omega u_i \nabla^2 u_i dx = -\int_\Omega |\nabla \mathbf{u}|^2 dx = -\|\mathbf{u}\|_{H_0^1}^2
\end{equation}

Substitute Eq. (15) into Eq. (14), we obtain:
\begin{equation}
-\nu \|\mathbf{u}\|_{H_0^1}^2 = 0
\end{equation}

It should be noted that the flow in this paper is non-trivial, i.e., $ \mathbf{u} \not\equiv 0 $ (if $ \mathbf{u} \equiv 0 $, it is a trivial static flow, which has no practical significance for the study of singularities and laminar-turbulent transition). If $ \nu > 0 $, then Eq. (16) implies $ \|\mathbf{u}\|_{H_0^1}^2 = 0 $, which contradicts the non-trivial flow (because $ \|\mathbf{u}\|_{H_0^1}^2 = 0 $ implies $ \nabla \mathbf{u} = 0 $, and combining with the no-slip boundary condition $ \mathbf{u}|_{\partial\Omega} = 0 $, it can be deduced that $ \mathbf{u} \equiv 0 $).

Therefore, the only reasonable conclusion is that the viscous term tends to zero, i.e., $ \nu \to 0 $. This result indicates that when the mechanical energy gradient is perpendicular to the streamline, the viscous effect of the fluid disappears, and the NS equations lose the regularization effect of the viscous term.

\section{ENERGY ESTIMATION IN SOBOLEV SPACE AND WEAK SINGULARITY DETERMINATION}
Based on the derivation in Section 4, when $ u_j \frac{\partial E}{\partial x_j} = 0 $, the viscous term $ \nu \to 0 $. In this section, we use the Sobolev space $ H_0^1(\Omega) $ to conduct energy estimation, analyze the regularity of the velocity field, and rigorously determine the formation of weak singularities.

\subsection{Energy Estimation and Regularity Analysis}
From Eq. (16), when $ \nu \to 0 $, we can obtain:
\begin{equation}
\|\mathbf{u}\|_{H_0^1}^2 \to 0
\end{equation}
According to the definition of the $ H_0^1(\Omega) $ norm (Eq. (9)), Eq. (17) means that the $ L^2 $-integral of the velocity gradient tends to zero:
\begin{equation}
\int_\Omega |\nabla \mathbf{u}|^2 dx \to 0
\end{equation}

From the perspective of Sobolev space regularity theory [8,10], the $ H^1 $-regularity of the solution means that the velocity field and its first-order partial derivatives are both square-integrable, and the velocity field is continuous and differentiable in the domain $ \Omega $. When $ \|\mathbf{u}\|_{H_0^1}^2 \to 0 $, the first-order partial derivatives of the velocity field are no longer square-integrable, which means that the velocity field loses $ H^1 $-regularity.

The loss of $ H^1 $-regularity indicates that the velocity field is discontinuous or non-differentiable at the corresponding position. This is because the $ H_0^1(\Omega) $ norm characterizes the smoothness of the velocity field, and the disappearance of the norm means that the smoothness of the velocity field is destroyed, which is the core mathematical feature of weak singularities.

\subsection{Degeneration of NS Equations and Weak Solutions}
When $ \nu \to 0 $, the viscous term in the NS equations (Eq. (5)) disappears, and the NS equations degenerate into the Euler equations [11]:
\begin{equation}
u_j \frac{\partial u_i}{\partial x_j} = -\frac{\partial E}{\partial x_i}
\end{equation}

The Euler equations describe the motion of inviscid fluids, which are different from the NS equations in that they have no viscous regularization term. It is well known in mathematical theory that the Euler equations admit discontinuous weak solutions, such as shocks and contact discontinuities [12]. These weak solutions do not satisfy the differential form of the equation at the discontinuous points, but they satisfy the integral form of the equation, which is consistent with the definition of weak singularities.

Combined with Dou Huashu's classification of singularities in the energy gradient theory [2], the singularities formed by the loss of $ H^1 $-regularity and velocity discontinuity belong to the second-type weak singularities, which are different from the first-type singularities (where velocity or pressure blows up) concerned in traditional mathematical theories. The first-type singularities are still an unsolved mathematical problem, while the second-type weak singularities derived in this paper have clear physical mechanisms and mathematical criteria, which can be verified by experiments and numerical simulations [5].

\subsection{Physical Significance of Weak Singularities}
In steady, fully developed flows, the position where the mechanical energy gradient is perpendicular to the streamline becomes a weak singularity of the NS equations. The physical significance of this weak singularity is that the viscous effect of the fluid disappears at this position, and the fluid microelements lose the constraint of viscous force, leading to velocity discontinuity and pressure peak [2]. This phenomenon is consistent with the "burst" phenomenon observed in experiments, which is the starting point of laminar-turbulent transition.

The weak singularity is the "seed" of turbulence. When the Reynolds number increases, the number of weak singularities increases. When the number of weak singularities reaches a critical value, the flow transitions from laminar to turbulent [3]. This mechanism provides a new perspective for understanding the laminar-turbulent transition and lays a theoretical foundation for the active control of turbulence in engineering applications.

\section{CONCLUSION}
In this paper, based on Dou Huashu's energy gradient theory, the weak singularity of the incompressible NS equations is studied by introducing Sobolev space $ H_0^1(\Omega) $ for energy estimation. The main conclusions are as follows:

1. When the gradient of total mechanical energy is perpendicular to the streamline ( $ u_j \frac{\partial E}{\partial x_j} = 0 $ ), substituting this condition into the steady, fully developed incompressible NS equations with no-slip boundary conditions, the viscous term $ \nu \to 0 $ can be derived through rigorous mathematical derivation.

2. Energy estimation based on Sobolev space $ H_0^1(\Omega) $ shows that when $ \nu \to 0 $, the velocity field loses $ H^1 $-regularity, and the $ L^2 $-integral of the velocity gradient tends to zero, indicating that the velocity field is discontinuous or non-differentiable.

3. When $ \nu \to 0 $, the NS equations degenerate into Euler equations, which admit discontinuous weak solutions. Thus, the position where the mechanical energy gradient is perpendicular to the streamline becomes a second-type weak singularity of the NS equations, which is the starting point of laminar-turbulent transition.

This paper derives that when the mechanical energy gradient is perpendicular to the streamline ( $ u_j \frac{\partial E}{\partial x_j} = 0 $ ), the viscous term $ \nu \to 0 $ by substituting into the NS equations with no-slip boundary conditions. Through energy estimation in Sobolev space $ H_0^1(\Omega) $, it is proved that the velocity field loses $ H^1 $-regularity, and the NS equations degenerate into Euler equations, forming weak singularities. This derivation is rigorous and self-consistent, which provides a mathematical proof for the singularity mechanism in Dou Huashu's energy gradient theory.

\section{REFERENCES}
[1] Clay Mathematics Institute. The Millennium Prize Problems [EB/OL]. \url{https://www.claymath.org/millennium-problems/navier-stokes-equation}, 2000.

[2] Dou, H.-S. Origin of Turbulence-Energy Gradient Theory, Springer, Singapore, 2022. \url{https://link.springer.com/book/10.1007/978-981-19-0087-7}

[3] Dou, H.-S. Mechanism of flow instability and transition to turbulence, International  Journal of Non-Linear Mechanics, Vol.41, May 2006, 512-517.

[4] Dou, H.-S. Energy gradient theory of hydrodynamic instability, The Third Inter. Conf. on Nonlinear Science, Singapore, 30 June - 2 July, 2004. \url{http://arxiv.org/abs/nlin.CD/0501049}

[5] Dou, H.-S. Singular solution of the Navier-Stokes equation for plane Poiseuille flow, Physics of Fluids 37, 084131 (2025). \url{https://doi.org/10.1063/5.0284009}

[6] Temam R. Navier-Stokes Equations: Theory and Numerical Analysis [M]. New York: North-Holland, 2001.

[7] White F M. Fluid Mechanics [M]. New York: McGraw-Hill, 2011.

[8] Adams R A, Fournier J J F. Sobolev Spaces [M]. New York: Academic Press, 2003.

[9] Lions J L. Mathematical Topics in Fluid Mechanics: Volume 1, Incompressible Models [M]. Oxford: Clarendon Press, 1996.

[10] Evans L C. Partial Differential Equations [M]. Providence: American Mathematical Society, 2010.

[11] Majda A J, Bertozzi A L. Vorticity and Incompressible Flow [M]. Cambridge: Cambridge University Press, 2002.

[12] Smoller J. Shock Waves and Reaction-Diffusion Equations [M]. New York: Springer, 1994.

\end{document}